\documentclass[technote,onecolumn]{IEEEtran}
\makeatletter
\def\ps@headings{%
\def\@oddhead{\mbox{}\scriptsize\rightmark \hfil \thepage}%
\def\@evenhead{\scriptsize\thepage \hfil \leftmark\mbox{}}%
\def\@oddfoot{}%
\def\@evenfoot{}}
\makeatother
\usepackage{cite}

\usepackage[pdftex]{graphicx}

\usepackage[cmex10]{amsmath}

\usepackage{url}

\usepackage{tikz}
\usetikzlibrary{arrows,snakes,backgrounds}
\tikzstyle{router}=[circle,draw=black!50,fill=black!20,thick]
\tikzstyle{funcnode}=[rectangle,draw=black!50,fill=black!20,thick]
\tikzstyle{varnode}=[circle,draw=black!50,fill=black!20,thick]

\makeatletter 
\newif\if@restonecol 
\makeatother

\usepackage[lined,linesnumbered,nofillcomment]{algorithm2e}

\begin{document}

\title{Real-Time Multi-path Tracking of Probabilistic Available Bandwidth}

\author{\IEEEauthorblockN{Frederic Thouin, Mark Coates, Michael Rabbat}
\IEEEauthorblockA{Department of Electrical and Computer Engineering\\
       McGill University, Montreal, Canada\\
       frederic.thouin@mail.mcgill.ca, mark.coates@mcgill.ca, michael.rabbat@mcgill.ca}}

\maketitle

\begin{abstract}
Applications such as traffic engineering and network provisioning can greatly benefit from knowing, in real time, what is the largest input rate at which it is possible to transmit on a given path without causing congestion.  We consider a probabilistic formulation for available bandwidth where the user specifies the probability of achieving an output rate almost as large as the input rate.  We are interested in estimating and tracking the network-wide probabilistic available bandwidth (PAB) on multiple paths simultaneously with minimal overhead on the network.  We propose a novel framework based on chirps, Bayesian inference, belief propagation and active sampling to estimate the PAB.  We also consider the time evolution of the PAB by forming a dynamic model and designing a tracking algorithm based on particle filters.  We implement our method in a lightweight and practical tool that has been deployed on the PlanetLab network to do online experiments.  We show through these experiments and simulations that our approach outperforms block-based algorithms in terms of input rate cost and probability of successful transmission.
\end{abstract}

\section{Introduction}

The latest research in overlay network routing~\cite{hir:07,lee:08} and anomaly detection~\cite{he:08anomaly} has shown that knowing the amount of available bandwidth (AB) of paths across the network can lead to better performance.  This knowledge could be helpful to many other applications, such as SLA compliance, network management, transport protocols, traffic engineering or admission control, but current available bandwidth estimation techniques and tools generally do not meet application-specific requirements in terms of accuracy, overhead, latency and reliability~\cite{gue:09applicability}.  Furthermore, the commonly-used definition of available bandwidth, in terms of utilization, and several tools and techniques suffer from three major deficiencies: 1) the models that associate the AB metric to the observed data are inaccurate in many practical scenarios; 2) the majority of existing tools produce a point estimate of average AB and do not provide a confidence interval; 3) most current tools are not well-adapted to multi-path or real-time estimation for applications such as traffic engineering and network provisioning~\cite{tho:10}.  

In this paper, we focus on the problem of tracking available bandwidth in real-time for multiple paths simultaneously.  Existing available bandwidth tracking tools, most of which are based on the Kalman filter~\cite{eke:06,ber:09,cab:08,yan:09,boz:09}, do not address the simultaneous tracking of the available bandwidths of multiple intersecting paths. The available bandwidth estimation techniques which do address multiple paths~\cite{hu:05,man:07bandwidth,yal:08} construct a single estimate for each path based on a block of data; they do not perform tracking.

In~\cite{tho:10}, we proposed a tool for block-based estimation of a metric (that we assumed to be static) called probabilistic available bandwidth (PAB).  
The PAB metric is defined to address the weaknesses of the utilization-based definition and related estimation tools and techniques\footnote{A precise definition of PAB is given in Sect.~\ref{sec:prob_statement}.}.  In this paper we allow PAB to vary with time and formulate inference as a filtering task.
Our main contribution is to propose a network-wide multi-path available bandwidth tracking procedure.  
We demonstrate a lightweight, practical implementation of the proposed approach that dramatically reduces the measurement overhead for PAB estimation and renders it feasible for multiple paths.
We extend the Bayesian formulation of our estimation problem and use Dynamic Bayesian Networks (DBNs)~\cite{gha:98,mur:02} to track the PAB.  
We use particle filtering (a method that combines sequential importance sampling~\cite{liu:98,dou:00} and resampling) to approximate the PAB of the links in the network with a mixture of weighted Gaussians.  
Our Bayesian approach allows us to use the marginal posteriors of the paths directly to produce confidence intervals.

The rest of the paper is organized as follows.  In Sect.~\ref{sec:related_work}, we summarize work related to available bandwidth tracking and network-wide estimation.  In Sect.~\ref{sec:prob_statement}, we provide the exact definition of probabilistic available bandwidth and formulate the tracking problem.  In Sect.~\ref{sec:measurement_model}, we present our measurement model constructed using chirps that provide considerable savings in terms of number of probes required compared to constant-rate packet trains.  In Sect.~\ref{sec:algorithm}, we outline the tracking methodology based on factor graphs, belief propagation and particle filtering.  In Sect.~\ref{sec:results}, we show the results of our simulation and online experiments on the PlanetLab network.  In Sect.~\ref{sec:conclusion}, we summarize our work and discuss future research possibilities.

\section{Related Work}
\label{sec:related_work}

For real-time estimation, the proposed techniques~\cite{eke:06,ber:09,cab:08,yan:09,boz:09} use Kalman filtering by taking advantage of the piecewise linear relation between utilization and available bandwidth.  The main drawback of the Kalman filter is that conditional probability distributions have to be Gaussian-linear.  Goldoni et al.\ use Vertical Horizontal Filtering that ignores sharp and non-persistent changes but quickly converges to the new value if they do persist~\cite{gol:09}.  However, since their tool was only tested in an environment with constant bit-rate cross-traffic, its performance for tracking is unknown.  

These tools cannot be applied directly to scenarios where the available bandwidths of multiple paths have to be simultaneously estimated.  The probes will generate interference on links shared by multiple paths, which can lead to significant underestimation, and also introduce an unacceptable overhead and overload on the network and the hosts~\cite{cro:09}.  Alternatively, each path can be probed independently in a sequence rather than simultaneously.  This approach is not only time-consuming but also very inefficient since it does not take advantage of the notable correlations in the AB when links are shared among paths.  The techniques that have been proposed for large scale scenarios do rely on the correlations between links or even between the various metric (route, number of hops, capacity) to reduce the number of probes required to produce accurate estimates~\cite{hu:05,man:07bandwidth,yal:08}.  However, all of them are limited to estimating and not tracking the available bandwidth.  Multi-path tracking has been proposed for other metrics; Coates and Nowak use sequential Monte Carlo inference to estimate and track internal delay characteristics~\cite{coa:02}.  

\section{Problem Statement}
\label{sec:prob_statement}

The probabilistic available bandwidth (PAB) of a path $p$ is defined in~\cite{tho:10} as the maximum input rate $r_p$ at which we can send traffic such that the output rate $r'_p$ is almost as large as  the input rate with specified probability\footnote{The probability is defined over all possible multi-packet flows of average rate equal to the input rate that can be transmitted during the measurement period.}.  
The PAB of each path $p$ is modelled as a discrete\footnote{We chose discrete, rather than continuous, random variables because it is not meaningful to have an infinite precision on the transmission rates.} random variable $y_p$.  The PABs of the constituent links are denoted by $x_{\ell}$.  The probability that the PAB of path $p$ is $r$ is then $\Pr(y_p = r)$.  For a $\epsilon > 0$ and $\gamma > 0$, we have:

\begin{equation}
y_p(t,\lambda) = \max \{r_p : \Pr(r'_p > r_p - \epsilon) \geq \gamma\}\label{eq:pab}
\end{equation}
where $y_p(t,\lambda)$ denotes the PAB at time $t$, for a measurement period $(t-\lambda,t]$, where $\lambda$ is the number of observations made in one measurement period\footnote{The average duration of one measurement using our tool is under 2 seconds.}.

The problem of interest in this paper is that of tracking PAB over time: for a user-specified $(\epsilon,\gamma)$ and known topology that consists of a set of $N$ links, denoted $\mathcal{L}$, and $M$ paths, denoted $\mathcal{P}$, we want to produce, at regular time intervals, estimates of the probabilistic available bandwidths for all paths in the network.  
Our measurement strategy is such that, within each measurement interval $(t-\lambda,t]$, we make $\lambda$ measurements; each one on a single path. 
Each measurement may probe one or several rates.  For each rate, we evaluate a binary outcome $z_t$ that indicates whether or not the output rate is within $\epsilon$ Mbps of the probing rate.  We use a binary outcome, rather than the output rate, because it is less sensitive to noisy measurements and it is also easier to construct an accurate likelihood function empirically.  
We want to identify the most informative measurement at each iteration $t$ such that, for successive windows of $\lambda$ measurements, we can compute $\Pr(y_p(t,\lambda) | z_{t-\lambda:t}$) for every path.  This marginal posterior can be used to produce credible intervals for the estimates, which can be employed to construct predictions of available bandwidth for subsequent time windows.

\subsection{Assumptions}

Our measurement and tracking methodology is based on three assumptions.  
\begin{enumerate}
\item Each link is modelled as a store-and-forward first-come first-serve router/switch.  If the network employs other forms of queueing or router-level Quality-of-Service provisioning, then we infer the PAB as seen by the class of packets transmitted as probes.
\item The routing topology of this network is known and it remains fixed during the tracking\footnote{Simulations in~\cite{tho:10} have shown the that effect of changes in the topology in terms of accuracy and convergence time is negligible.}.  The IP addresses of the routers and the mapping to a physical topology is done using $\texttt{traceroute}$, which provides sufficiently accurate topology estimates for us to assess the performance of our algorithm.  If there is per-packet load balancing, only one of the paths will be identified.  Destination-based load balancing does not affect our method.
\item There is a single tight link on each path that determines the PAB of that path.  The PAB of a path is the minimum of the PABs of its constituent links.  The presence of more than one tight link results in noise propagated in the factor graph during the estimation procedure.
\end{enumerate}

\section{Measurement Model}
\label{sec:measurement_model}

One measurement methodology to estimate PAB consists of sending constant-rate trains of packets at probing rate $r_p$ and measuring the receiving rate $r^{'}_p$~\cite{car:96,dov:01,jai:03,tho:10}.  The result of each train is a binary decision as to whether the probing rate was greater or smaller than the receiving rate.  Estimating the receiving rate with sufficient accuracy requires a minimal number of packets.  This means that for each rate probed, there is a significant overhead (load on the network) involved.   To reduce this overhead, an alternative solution to constant-rate packet trains is chirps~\cite{rib:00,rib:03,gol:09}.  A chirp train consists of sending a train of packets with exponentially increasing packet spacing.  By varying the spacing between the packets rather than keeping it constant, it is possible to probe multiple rates with a single chirp.

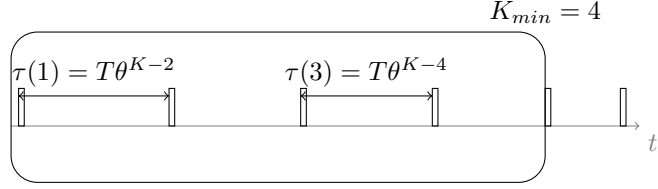
\begin{figure}[!h]
	\centering
	\begin{tikzpicture}
		\foreach \y in {0}{
			\draw[help lines,->] (-1.1,\y) -- (7.25,\y) node[anchor=north west] {$t$};
			\foreach \x in {-1,1,2.75,4.5,6,7}
				\draw (\x,\y) rectangle (\x+0.075,\y+0.5);}
		\draw [rounded corners=10pt] (-1.1,-0.75) rectangle (6,1.25) node[anchor=south] {$K_{min}=4$}; 
		\draw[<->] (-1,0.40) -- (1,0.40) node [midway,anchor=south] {$\tau(1) = T\theta^{K-2}$};
		\draw[<->] (2.75,0.40) -- (4.5,0.40) node [midway,anchor=south] {$\tau(3) = T\theta^{K-4}$};
	\end{tikzpicture}
	\caption{Chirp of $K=6$ packets.  Probing rate with sliding window of $K_{min} = 4$ packets of $P=1$KB each.  This chirp probes $K-K_{min}=4$ rates.\label{fig:chirp}}
\end{figure}

A chirp is defined by its length $K$, window size $K_{min}$, spacing factor $\theta$, minimum spacing $T$ and packet size $P$ (Fig.~\ref{fig:chirp}).
The length $K$ represents the total number of packets in the chirp.  Increasing the value of $K$ allows for a larger number of rates to be probed with a single chirp, but it also increases the number of bytes injected into the network.  We do not want our chirp to be too intrusive, but we are interested in having the ability to cover a wide enough range of rates with a single chirp (potentially the entire range of possible values).  

To probe multiple rates with a single chirp, we apply a sliding window of size $K_{min}$ across the chirp.
The size of the sliding window represents a trade-off between accuracy and range of rates.  
The maximum number of rates probed by one chirp is $K'=K-K_{min}$, which means that reducing the value of $K_{min}$ will allow for a larger range of rates to be probed.
However, it will also result in noisier measurements.  
In pathChirp~\cite{rib:03}, a single packet pair is used to estimate one probing rate ($K_{min} = 1$).  In the constant-rate approach, all the packets in the train are used ($K_{min} = K-1$).  We interpret each window of packets as a single probe with an input rate equal to the average rate over that window.  
The average input rate of the $k$th window in the chirp is calculated as follows:
\begin{equation}r_p(k) = \frac{K_{min} \cdot P}{\sum_{i = k}^ {k + K_{min} - 1} \tau(i)}
\end{equation}
where $\tau(i)$ is the spacing between the $i$th and $i+1$th packet, i.e., the time elapsed between the departure of the two packets.  


The spacing factor $\theta$ and the minimum spacing $T$ determine the packet spacing:
\begin{equation}\tau(i) = T\theta^{K-(i+1)}  \quad \forall{i}=1..K-1
\end{equation}
For a fixed packet size $P$, these values are adjusted to obtain the desired range of probing rates $[r_p(1), r_p(K')]$ covered by the chirp.  
In the constant-rate method, the spacing is constant throughout the entire train of packets, which results in a single probing rate.  
The main feature of chirps is that the spacing varies in order to test multiple probing rates with a single train.  
We fix the value of $T$ and $\theta$, by satisfying the following two equalities:
\begin{equation}\frac{K_{min} \cdot P}{\sum_{i = 1}^ {K_{min}} \tau(i)} = r_p(1) \qquad \frac{K_{min} \cdot P}{\sum_{i = K-K_{min}} ^ {K-1} \tau(i)} = r_p(K')
\end{equation}
where $r_p(1)$ is the smallest rate we wish to probe and $r_p(K')$ is the largest.
We note that a chirp does not necessarily probe $K'$ different rates.  As the range of rates gets smaller, it is possible that more than one window of packets will probe the same rate.      Although we could reduce the chirp size to avoid this situation, we keep $K$ constant and use the extra measurements to reduce the amount of noise and make the outcome more reliable.

Each chirp provides up to $K'=K-K_{min}$ measurements if no packets are dropped or discarded\footnote{A packet is discarded if the time elapsed between the departure of the previous packet and this one is greater than $\tau(i)$, which means that there was an unusual delay at the sender.  A discarded packet is simply excluded from the vector of delays $\tau'$.}.  Each window of $K_{min}$ packets is interpreted as a single probe rate at the average input rate over that window.  We then measure the output rate $r_p'$ for the same set of packets:
\begin{equation}
r_p'(k) = \frac{K_{min} \cdot P}{\sum_{i = k}^ {k + K_{min} - 1} \tau'(i)}
\end{equation}
where $\tau'(i)$ is the delay between the arrival of the $i$th and $i+1$th packet at the receiver.  Using the input and output rate vectors, we can compute a vector of binary outcomes:
\begin{equation}
z(k) = \mathbf{1}(r_p' \geq r_p - \epsilon) \quad \forall k=1..K'
\end{equation}
where $\mathbf{1}(\cdot)$ is the indicator function.  

According to our simulations and online experiments, the chirp-based approach provides significant savings in terms of time required to form reliable estimates (at least $70\%$) and number of measurements required for accurate block-based estimation (at least $88\%$).  Whereas the number of measurements increases drastically when using constant-rate packet trains for values of $\gamma$ close to one, there is almost no impact when using chirps.

\subsection{Likelihood function}
\label{ssec:likelihood}

To use the outcome of a chirp measurement in our Bayesian framework, we need to construct a likelihood function $L(\mathbf{z} | y_p)$ where $\mathbf{z} = [z(1) \dots z(K')]$ and $y_p$ is the PAB of the probed path.  To simplify the learning of such a function, we first assume that the outcomes from the different probed rates within a single chirp are independent:
\begin{equation}
L(\mathbf{z} | y_p) = \prod_{i=1}^{K'} L(z(i) | y_p)
\end{equation}

As an example, we learn $L(z|y_p)$ empirically from 71050 data points collected over a 24 hour period from 42 different paths for $K = 75$, $K_{min} = 15$ and $\epsilon=5$.  As a comparison, for a constant-rate approach with three trains of 25 packets for each rate, the chirp equivalent (in terms of bytes per measurement) is $K=75$ and $K_{min} = 25$.  For the same number of packets, a chirp can probe 50 rates instead of 3.  In this case, we choose $K_{min} = 15$ because we did not notice a significant difference in the amount of noise and we are able to probe up to 10 extra rates.  


For our likelihood function, we choose the form $L(z=1|r_p,y_p) = {\mathrm{logsig}}(\alpha \cdot (r_p - y_p))$.  Here $\alpha$ is a constant learned empirically co-jointly with the estimate of $y_p$ through a single regression procedure.   The best fit in this regression is determined by minimizing the mean-squared error.  
The regression identifies $\alpha = -0.27$ with a MSE of $0.08$ over the range from $[1,100]$ Mbps\footnote{We omit a graphical depiction of the functional regression results due to lack of space.}.
In theory, the learning procedure should be repeated every time we estimate a new set of paths.  However, from the data collected during multiple experiments, the sigmoid function is consistently a good fit and the value of $\alpha$ rarely changes significantly.  It is also interesting to note that the value of $\alpha$ is almost the same as the one obtained for the learning procedure in~\cite{tho:10} for a completely different set of paths and measurement methodology (chirps instead of constant-rate packet trains).  This suggests that the behaviour of a single window in a chirp is similar to a constant-rate packet train (which validates in part our independence assumption) and that this function is not specific to a single topology.

\subsection{Active Sampling}
\label{ssec:active_sampling}
Before making each measurement, we must choose which path to probe and the range of rates of the chirp. Adaptive selection algorithms, which consist of using information collected with previous measurements to make decisions about the future, can provide important reductions in the number of probes~\cite{cas:07}.

We employ a probabilistic greedy active learning procedure to select the path to probe at each iteration. The procedure assigns a selection probability to each path that is proportional to the width of the current confidence interval of the path's PAB; it then chooses a path at random according to the assigned probabilities. This means that paths are more likely to be probed if there is more uncertainty about their PABs, and the range of rates probed is mostly likely to contain the PAB of the particular path, given all previous measurements.

From the marginal posterior of each path, we can calculate a confidence range where a specified fraction of the probability mass lies. We use the bounds of the confidence range as the range of rates for the next chirp. For a pre-specified sliding window size, we obtain the desired range of input rates by adjusting the packet spacing and the length of the chirp.

\section{Tracking Model and Algorithm}
\label{sec:algorithm}
We track the PAB by forming a dynamic model for how it evolves over time.  
This allows us to implement a forgetting factor; the impact of the first measurement is gradually eliminated because the dynamic model incorporates diffusion.
The model we adopt is a dynamic Bayesian network (DBN), which consists of two temporal ``slices'' related by a transition function~\cite{mur:02,gha:98}.
Each slice contains variables interconnected based on their dependencies, including a set of variables used for the distributed representation of the (unknown) belief state; in our case, the PABs.  The transition function determines the evolution of the variables between the slices; it is a model for the dynamics of the PABs.  In this section, we describe this model in detail and describe the algorithm used to track and estimate the path PAB.

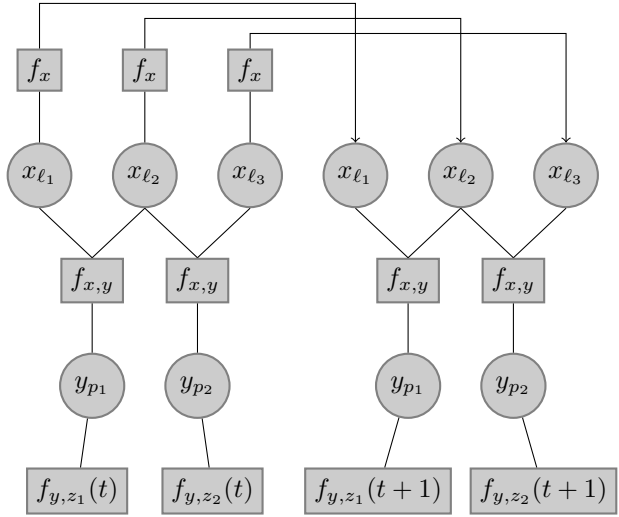
\begin{figure}[!h]
	\centering
	\begin{tikzpicture}[node distance=1.4cm]
		\node[varnode] (x1) {$x_{\ell_1}$};
		\node[varnode] (x2) [right of=x1] {$x_{\ell_2}$};
		\node[varnode] (x3) [right of=x2] {$x_{\ell_3}$};
		\node[funcnode] (fx1) [above of=x1] {$f_x$};		
		\node[funcnode] (fx2) [above of=x2] {$f_x$};
		\node[funcnode] (fx3) [above of=x3] {$f_x$};		
		
		\draw (x1.north) -- (fx1.south);
		\draw (x2.north) -- (fx2.south);
		\draw (x3.north) -- (fx3.south);
		
		\node[funcnode] (fy1) [below of=x1,xshift=7mm] {$f_{x,y}$};
		\node[funcnode] (fy2) [below of=x2,xshift=7mm] {$f_{x,y}$};
		
		\node[varnode] (y1) [below of=fy1] {$y_{p_1}$};
		\node[varnode] (y2) [below of=fy2] {$y_{p_2}$};
		\draw (y1) -- (fy1);
		\draw (y2) -- (fy2);
		
		\draw (fy1.north) -- (x1.south);
		\draw (fy1.north) -- (x2.south);
		
		\draw (fy2.north) -- (x2.south);
		\draw (fy2.north) -- (x3.south);
		
		\node[funcnode] (fz1) [below of=y1,xshift=-2mm] {$f_{y,z_1}(t)$};		
		\draw (y1) -- (fz1);					
		\node[funcnode] (fz2) [below of=y2,xshift=2mm] {$f_{y,z_2}(t)$};		
		\draw (y2) -- (fz2);	
		
		\node[varnode] (x1t2) [right of=x3] {$x_{\ell_1}$};
		\node[varnode] (x2t2) [right of=x1t2] {$x_{\ell_2}$};
		\node[varnode] (x3t2) [right of=x2t2] {$x_{\ell_3}$};	
		
		\draw[->] (fx1.north) -- ++(0,0.6) -| (x1t2.north);
		\draw[->] (fx2.north) -- ++(0,0.4) -| (x2t2.north);
		\draw[->] (fx3.north) -- ++(0,0.2) -| (x3t2.north);				
		
		\node[funcnode] (fy1t2) [below of=x1t2,xshift=7mm] {$f_{x,y}$};
		\node[funcnode] (fy2t2) [below of=x2t2,xshift=7mm] {$f_{x,y}$};
		
		\node[varnode] (y1t2) [below of=fy1t2] {$y_{p_1}$};
		\node[varnode] (y2t2) [below of=fy2t2] {$y_{p_2}$};
		\draw (y1t2) -- (fy1t2);
		\draw (y2t2) -- (fy2t2);
		
		\draw (fy1t2.north) -- (x1t2.south);
		\draw (fy1t2.north) -- (x2t2.south);
		
		\draw (fy2t2.north) -- (x2t2.south);
		\draw (fy2t2.north) -- (x3t2.south);
		
		\node[funcnode] (fz1t2) [below of=y1t2,xshift=-4mm] {$f_{y,z_1}(t+1)$};		
		\draw (y1t2) -- (fz1t2);					
		\node[funcnode] (fz2t2) [below of=y2t2,xshift=4mm] {$f_{y,z_2}(t+1)$};		
		\draw (y2t2) -- (fz2t2);

	\end{tikzpicture}
	\caption{Two slices of the DBN.  Circles represent random variables; PAB of links and path.  Rectangles are factors in the joint distribution.  $f_x$ is the transition function between the two slices of the DBN.\label{fig:DBN}}
\end{figure}
\subsection{Bayesian Inference}

Each slice of the DBN is a factor graph; a graphical representation of the factorization of the joint distribution from which we calculate the marginal posteriors~\cite{pea:88}.
The factor graph is composed of nodes for each variable and factor (in the distribution of interest) connected through edges based on their dependencies.  In our case, the variables consist of the PAB of every path ($y_p$) and every link ($x_{\ell}$).  The first factor, $f_{x,y}$, represents the relation between the PAB of links and paths; the PAB of a path is equal to the minimum of the PABs of all of its constituent links.  For example, in Fig.~\ref{fig:DBN}, path $p_1$ is constituted of links $\ell_1$ and $\ell_2$.  In the Bayesian inference framework, the posterior is proportional to the product of the likelihood and the prior distribution.  For each path, there is a likelihood function $f_{y,z}$ equal to the product of all the measurement outcomes on that path during the measurement period.  After each measurement, we update the likelihood of the measured path (the existing likelihood is multiplied with the likelihood of the last measurement) and we run the belief propagation algorithm on the factor graph.  This allows us to compute the marginal posterior of each path efficiently in order to determine the next most informative measurement (according to the algorithm presented in Sect.~\ref{ssec:active_sampling}).  The only variables in the graph that depend on their value in the previous iteration are the variables that describe the belief state; i.e., the $x_{\ell}$s.  The prior distribution of each link is used as a transition function.

\subsection{Belief State}

The belief state is represented by the set of random variables $x_{\ell}$, i.e., the link PABs (from which we can calculate all the path PABs).  
We approximate the belief state with a weighted mixture of Gaussians for each link~\cite{kot:03} as follows:

\begin{equation}
\widehat{\Pr}(x_{\ell}(t)|\mu_{\ell}^v(t)) = \mathcal{N}(\mu_{\ell}^v(t),\sigma_{\mu})\label{eq:mu_v}
\end{equation}

\begin{equation}
\widehat{\Pr}(x_{\ell}(t)|\mu_{\ell}(t)) = \sum_{v = 1}^{N_v} w_{\ell}^v(t) \widehat{\Pr}(x_{\ell}(t)|\mu_{\ell}^v(t))\label{eq:mu}
\end{equation}
where $\mu_{\ell}(t) = [\mu_{\ell}^1(t),\dots,\mu_{\ell}^{N_v}(t)]$ is the vector of means of the Gaussians of link $\ell$ at time $t$, $w_{\ell}^v(t)$ is the weight of each Gaussian in the mixture, $\mathcal{N}(\mu,\sigma)$ is a normal distribution with mean $\mu$ and variance $\sigma$.  Here, $\sigma_{\mu}$ is taken to be a constant and is equal to 1.  Initially, the means of the Gaussians are drawn from a uniform distribution that covers the range of possible PAB, $[B_{min},B_{max}]$; $\mu_{\ell}^v \sim \mathcal{U}(B_{min},B_{max})$ and weights are all equal; $w_{\ell}^v = 1/N_v$.  Sequential importance sampling~\cite{liu:98,dou:00} is employed to update the means and weights of the Gaussians each time the belief is propagated to a new measurement period.  Ideally, we would use a $N$-dimensional weighted Gaussian mixture to approximate the joint distribution of the PAB of each link; $\Pr(x_{\ell_1},\dots,x_{\ell_N})$, rather than individual mixtures for each link.  However, such a large space becomes intractable and the particle filter is known to often fail for high-dimensional problems.  Therefore, we consider each link independently (each link has its own weighted Gaussian mixture) and approximate the joint distribution as follows:

\begin{equation}
\Pr(x_1(t),\dots,x_N(t) | z_{1:t}) \approx \prod_{\ell = 1}^N \Pr(x_{\ell}(t) | z_{1:t}).
\end{equation}
The result of this approximation is that we are propagating the marginals of the links instead of the joint prior in the factor graph.  
However, the belief propagation algorithm still operates on the joint prior even if it is producing marginals.  
The approximation significantly decreases the dimensionality and complexity of the problem.  
In introduces an approximation error, but our simulations and experiments indicate that the approximation is reasonable and leads to acceptable tracking performance.

\subsection{Transition Function}

Every time we gather $\lambda$ observations, we perform the transition between the two slices of the DBN.
The means of the Gaussians are propagated to the next slice of the DBN as follows:

\begin{equation}
\mu_{\ell}^v(t+1) = \mu_{\ell}^v(t) + \epsilon_h\label{eq:mu_t}
\end{equation}
where $\epsilon_h$ is sampled from a Gaussian distribution $\mathcal{N}(0,\sigma_h)$.  

A product of marginals will produce a more diffuse distribution than the original joint distribution. So by adopting the approximation, we already introduce diffusion in the joint posterior (the new prior).  For that reason, we downplay the true variance in the temporal dynamics when we choose the value of $\sigma_h$.  For all our experiments, we use $\sigma_h = 4$ based on data collected during our online experimentation.

\subsection{Sequential Importance Sampling and Resampling}

Once the transition between the two slices is complete, before taking new measurements, we update the weights of the mixtures based on the observations gathered in the previous slice.  In the context of a DBN, this procedure is called likelihood weighting~\cite{fun:89,sha:89} and consists of taking the product of all the observed nodes (for a given link, all the observed paths that include this link).

Let $z_p^{\lambda}=z_p(t-\lambda+1,t)$ be the observations on path $p$ during the last interval of $\lambda$ measurements, $P_{\ell}$ be the set of paths that include link $\ell$ and $P_{\ell}^{'}$ the subset of $P_{\ell}$ for which $z_p^{\lambda}$ is not empty (i.e., the set of paths that include link $\ell$ observed during the last $\lambda$ measurements).  Then, the weights are updated as follows: 

\begin{equation}
\widetilde{w}_{\ell}^v(t+1) = w_{\ell}^v(t) \prod_{p \in \Pr_{\ell}^{'}} \widetilde{w}_{\ell,p}^v(t)\label{eq:w}
\end{equation}

\begin{eqnarray}
\widetilde{w}_{\ell,p}^v(t) &=& \Pr(z_p^{\tau} | \mu (t))\nonumber\\
&=& \sum_{x_{\ell} \in L_p} \left[ \Pr(z_p^{\tau} | x_{\ell}(t)) \widehat{\Pr}(x_{\ell}(t)|\mu_{\ell}^v(t)) \cdot \prod_{l \in L_p \setminus \ell} \widehat{\Pr}(x_{l}(t)|\mu_{l}(t))\right]\label{eq:w2}
\end{eqnarray}
where $L_p$ is the set of links composing path $p$.  The factorization in the joint distribution reduces the complexity of the weight computation.  
Even if the expression still looks computationally expensive, it reduces, for each link in $L_p$ to the multiplication of two $B\times B$ matrices, where $B = B_{max} - B_{min} + 1$ is number of possible discrete values for the PAB.  The weights are then normalized as follows:

\begin{equation}
w_{\ell}^v(t) = \widetilde{w}_{\ell}^v(t) / \sum_{v=1}^{N_v} \widetilde{w}_{\ell}^v(t)\label{eq:w_norm}
\end{equation}

Because of the multiplicative update applied to each weight in (\ref{eq:w}), the importance of some weights may converge towards zero, which effectively makes the associated Gaussian useless and also increases the variance of the estimator.  To address this issue, if the number of effective Gaussians, $N_{eff}^{\ell}$, drops below a specified threshold, the means are resampled.  The effective number of Gaussians is calculated as follows:

\begin{equation}
	N_{eff}^{\ell} = \frac{1}{\sum_{v = 1}^{N_p} {w^v_{\ell}}^2}
\end{equation}

We use a form of resampling called importance resampling~\cite{dou:00}, which consists of drawing a new set of $N_v$ samples from a multinomial distribution with parameters $N_p$ and the set of current weights $w^v_{\ell}$.  The new weights after this procedure are $1/N_v$.

\subsection{Algorithm}

The procedure we described in this section is summarized in Alg.~\ref{alg:tracking}.  Lines 5-12 represent the measurement methodology described in Sect.~\ref{sec:measurement_model} and the belief propagation.  The transition between the two-slices of the DBN occurs at line 13.  Lines 14-28 include the particle filtering, the sequential importance sampling and resampling.

\begin{algorithm}

initialize particles $\mu \sim \mathcal{U}(B_{min},B_{max})$\;
create DBN\;
$t = 1$\;
\While{tracking}
{
	\Repeat{$t\mod\lambda = 0$}
	{
		select path for next measurement\;
		select rates for next measurement\;
		make measurement\;
		run belief propagation\;
		compute confidence intervals\;		
		$t = t + 1$
	}

	propagate particles using (\ref{eq:mu_t})\;
		
	\ForEach {link $\ell \in \mathcal{L}$}
	{
		\ForEach {particle $v$}
		{
			\ForEach{observed path $p \in P_{\ell}$}
			{
				compute partial weights $w_{\ell,p}^v$ using (\ref{eq:w2})
			}
		}
				
		\eIf {link $\ell$ is observed}
		{
			update weights using (\ref{eq:w}) and (\ref{eq:w_norm})\;
			resample particles (if necessary)\;
			update prior using (\ref{eq:mu_v}) and (\ref{eq:mu})\;
		}{
			$w_{\ell}^v(t+1) = w_{\ell}^v(t)$
		}
		
	}	
}
\caption{Belief Propagation Particle Filtering (BP-PF) tracking algorithm\label{alg:tracking}}
\end{algorithm}

\section{Experiments and Results}
\label{sec:results}

\subsection{Matlab Simulations}

To assess the performance of our PF-based algorithm, we simulate a network environment where the PAB varies at every time instant.  We use a topology extracted from the PlanetLab network that consists of 72 paths (all possible paths between 9 end nodes) and 134 links.  The PAB of each link is initially drawn from a uniform distribution in the range $[B_{min}=1 Mbps,B_{max}=100 Mbps]$.  The PAB of each path is obtained from the minimum of the PAB of all its constituent links and is defined as $y_p = \max_{r_p}\Pr\{r'_p \geq r_p - 5\} \geq 0.8$ ($\epsilon = 5$ and $\gamma = 0.8$).  At every instant $t$, a measurement is taken.  The outcome $z$ is generated based on the likelihood model presented in Sect.~\ref{ssec:likelihood}.  The PAB of each link also changes at every instant $t$ according the probabilities\footnote{The probabilities correspond to a binomial distribution shifted by 2; $\mathcal{B}(x+2; N=4,p=0.5)$.} shown in Tab.~\ref{tab:prob_link_change}.  The PAB are generated for $T=1000$ measurements; $1 \leq t \leq  1000$.

\begin{table}[!h]
\caption{Probabilities of link PAB variation $\delta_{\ell}$: $x_{\ell}(t+1) = x_{\ell}(t) + \delta_{\ell}$.\label{tab:prob_link_change}}
\centering
\begin{tabular}{c c c c c}
$p(\delta_{\ell} = -2)$ & $p(\delta_{\ell} = -1)$ & $p(\delta_{\ell} = 0)$ & $p(\delta_{\ell} = 1)$ & $p(\delta_{\ell} = 2)$\\
\hline
$0.0625$ & $0.2500$ & $0.3750$ & $0.2500$ & $0.0625$
\end{tabular}
\end{table}

Our algorithm produces confidence intervals for each path at every $\lambda$ measurements, which are used to produce estimates of the PAB, $\hat{y}_p(t)$.  For these simulations, the prediction is valid for the upcoming interval of $\lambda=10$ measurements\footnote{In practice $\lambda=10$ is the equivalent of a 20 seconds interval.  The duration of this time interval must be chosen carefully such that it is long enough to gather sufficient information from measurements to update the PF accurately and short enough to make sure that the predicted values do not become outdated too quickly.  From our simulations and online experiments, $\lambda=10$ is a good compromise, but it is important to mention that this value can be adjusted by the operator and is totally dependent on the network dynamics and the application requirements.}; $[t,t+10]$.  The confidence interval is obtained directly from the marginal posterior of the path in two possible ways: i) smallest confidence interval that includes at least $\eta$ of the probability mass or ii) confidence interval of size $\beta$ that has the most probability mass (confidence level).  We use the former approach with $\eta = 0.95$.  For these simulations, we compare three possible ways to select $\hat{y}_p(t)$: i) lower bound, ii) 25th percentile and iii) median of the confidence interval\footnote{To determine the 25th percentile and median values, we first normalize the probabilities so that the sum of probabilities in the confidence interval is equal to 1.}.

From the definition of the PAB and our problem statement, we are interested in determining the largest input rate such that the output rate is within $\epsilon$ of the input rate with probability $\gamma$.  From our likelihood model, any rate smaller or equal to the PAB satisfies that constraint.  Such a rate is not necessarily optimal, but it represents a feasible solution; underestimates are preferable to overestimates.  To measure the performance of our approach we define three different metrics.  First, we evaluate the probability of identifying a feasible solution for path $p$, $PS_p(t)$, as a running average up to time $t$:  

\begin{equation}
PS_p(t) = \frac{1}{t} \sum_{k = 1}^{t} \mathbf{1}(\hat{y}_p(k) \leq y_p(k)) \label{eq:cong_avoid}
\end{equation}

For any feasible solution, the PAB was underestimated by a certain margin that represents the cost we are paying in terms of input rate.  In other words, by how much the input rate could have been increased without reducing the probability of successful transmission below $\gamma$.  At any instant $t$ where the estimate for path $p$ is smaller of equal to the PAB, we can calculate the cost in rate, $CR_p(t)$, as follows:

\begin{equation}
CR_p(t) = \hat{y}_p(t) - y_p(t) \label{eq:rate_diff}
\end{equation}

For estimates greater than the PAB, there is no penalty in terms of input rate, but there is one in terms of probability of success.  From the likelihood model, any rate greater than the PAB has $L(z=1) < \gamma$.  For a path $p$ where the estimate at time $t$ is greater than the PAB, we measure the cost paid in probability of success, $CP_p(t)$, as follows:

\begin{equation}
CP_p(t) = L(z=1|\hat{y_p},y_p) - \gamma \label{eq:cong_diff}
\end{equation}

For these three metrics, we compare our PF-based algorithm with two other block-based algorithms.  The first one, BB, is similar to our previous approach~\cite{tho:10} and assumes that the PAB is constant during the entire estimation procedure.  In that context, every measurement has the same weight and the estimate produced at time $t$ uses every observation from $z_{1:t}$.  This algorithm is not designed for tracking, but rather to produce confidence intervals for each path that satisfy tightness criteria as fast as possible.  The second algorithm, BB-R, is a variation of BB that only uses the last $\lambda$ measurements to produce the estimates.  It basically consists of re-running BB at every $\lambda$ measurements.  This approach discards all the information obtained from previous observations and therefore produces much wider confidence intervals, but reacts more quickly to changes in the system.  For these simulations, we used mixtures of $N_v = 100$ Gaussians for each link and a threshold of $N_{eff}=10$ for resampling.  We also simulated for $N_v = [50,250,500,1000]$ and $N_{eff} = [5,15]$, but did not observe any significant variation on the results.

\begin{figure}[!h]
	\centering
	\includegraphics[width=0.4\linewidth]{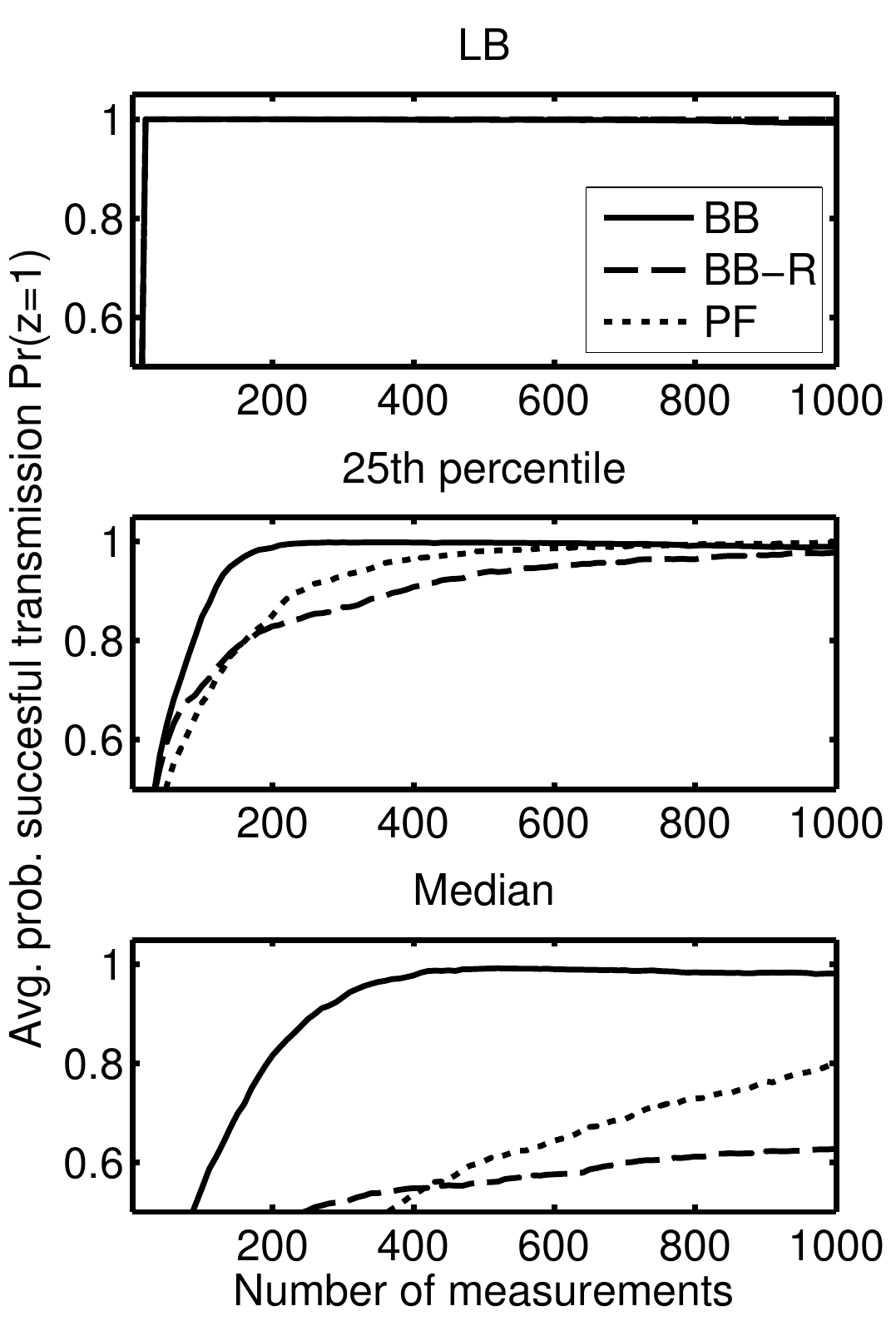}
	\caption{Running average probability of successful transmission as a function of the number of measurements.  Probability is averaged over 72 paths and 30 simulations.  Three selection modes for estimate of PAB $\hat{y}_p(t)$: the lower bound of the confidence interval (TOP), 25th percentile of the confidence interval (MIDDLE) and median of the confidence interval (BOTTOM).\label{fig:cong_avoid}}
\end{figure}

In Fig.~\ref{fig:cong_avoid}, we show the evolution as a function of the number of measurements of the probability of successful transmission for the three proposed methods.  The value at $t$ is first averaged over all time instants and then over all 72 paths.  The experiment was repeated 30 times with different link PAB initial values.  
For the first two selection methods (lower bound, 25th percentile, median), all techniques are able to avoid congestion almost all the time.  However, when probing at the median, only the BB approach is able to avoid congestion; our PF-based approach drops to $80\%$ congestion avoidance whereas the BB-R approach drops to almost $60\%$.
This metric is not really significant by itself because it is easy to avoid congestion if the estimate is conservative and much lower than the PAB.  This is the reason why we also look into the cost in terms of input rate and percentage of success.

\begin{figure}[!h]
	\centering
	\includegraphics[width= 0.4\linewidth]{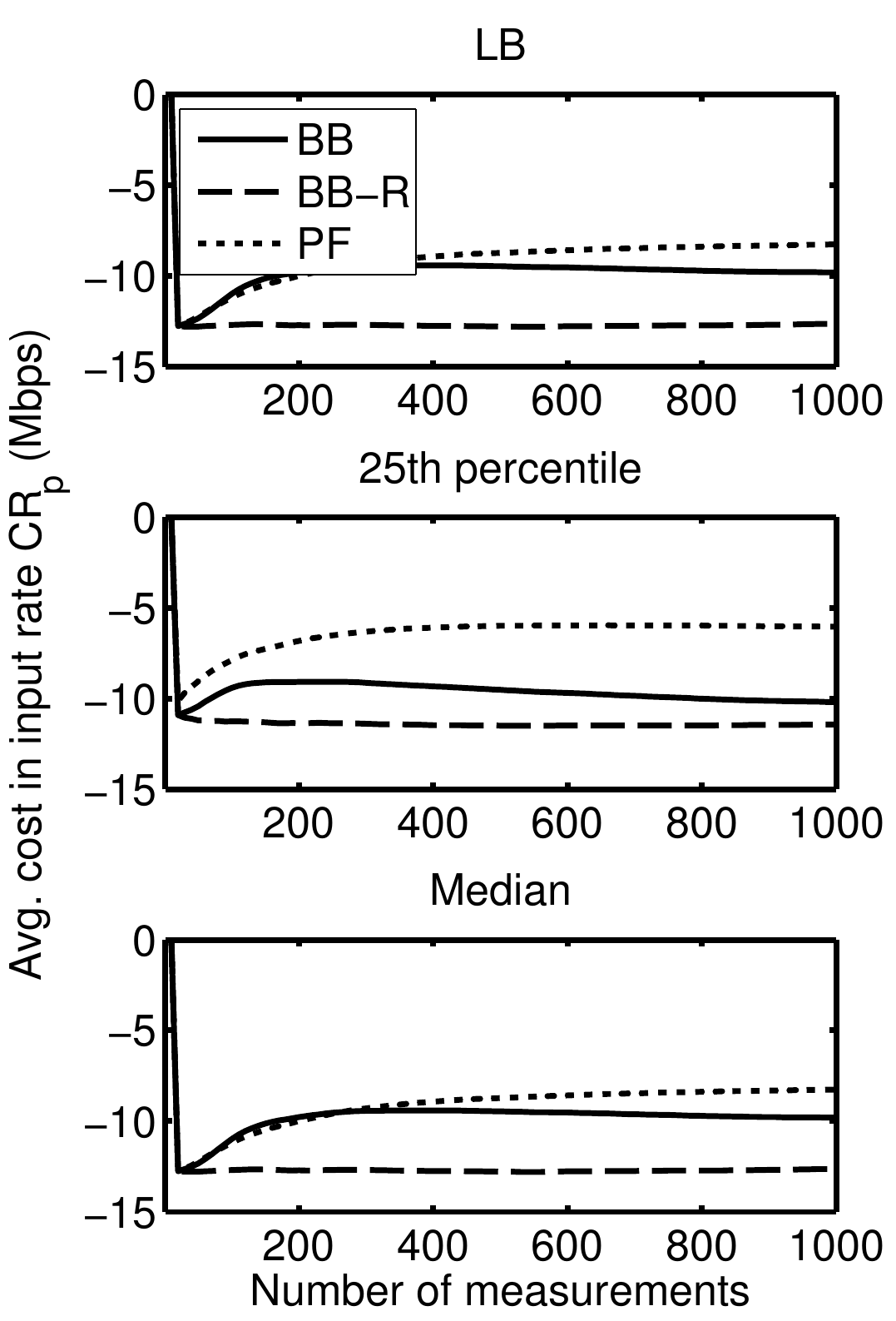}
	\caption{Cost in terms of input rate when estimate is smaller than the PAB.  Running average of the difference between the input rate and PAB.  The difference is averaged over 72 paths and 30 simulations.  Three selection modes for estimate of PAB $\hat{y}_p(t)$: the lower bound of the confidence interval (TOP), 25th percentile of the confidence interval (MIDDLE) and median of the confidence interval (BOTTOM).\label{fig:rate_diff}}
\end{figure}

In Fig.~\ref{fig:rate_diff}, we show the cost in terms of input rate for estimates smaller than the PAB.  The values represent running averages over all 72 paths and time instants, i.e., over $M\times{t}$ values at time $t$.  The experiment was repeated 30 times with different link PAB initial values and the running averages are averaged over all these experiments.
The PF outperforms the block-based approaches for all three estimation choices.  The cost decreases when the input rate is chosen to be around the 25th percentile to around 5 Mbps, which is 5Mbps closer than other algorithms.
The BB always underestimates the PAB by 10 Mbps.  This can be explained by the fact that the confidence intervals become very tight after a few hundred measurements and there is not much difference between the lower bound and the median.  
On the other hand, the BB-R algorithm always underestimates the PAB significantly.  

\begin{figure}[!h]
	\centering
	\includegraphics[width= 0.4\linewidth]{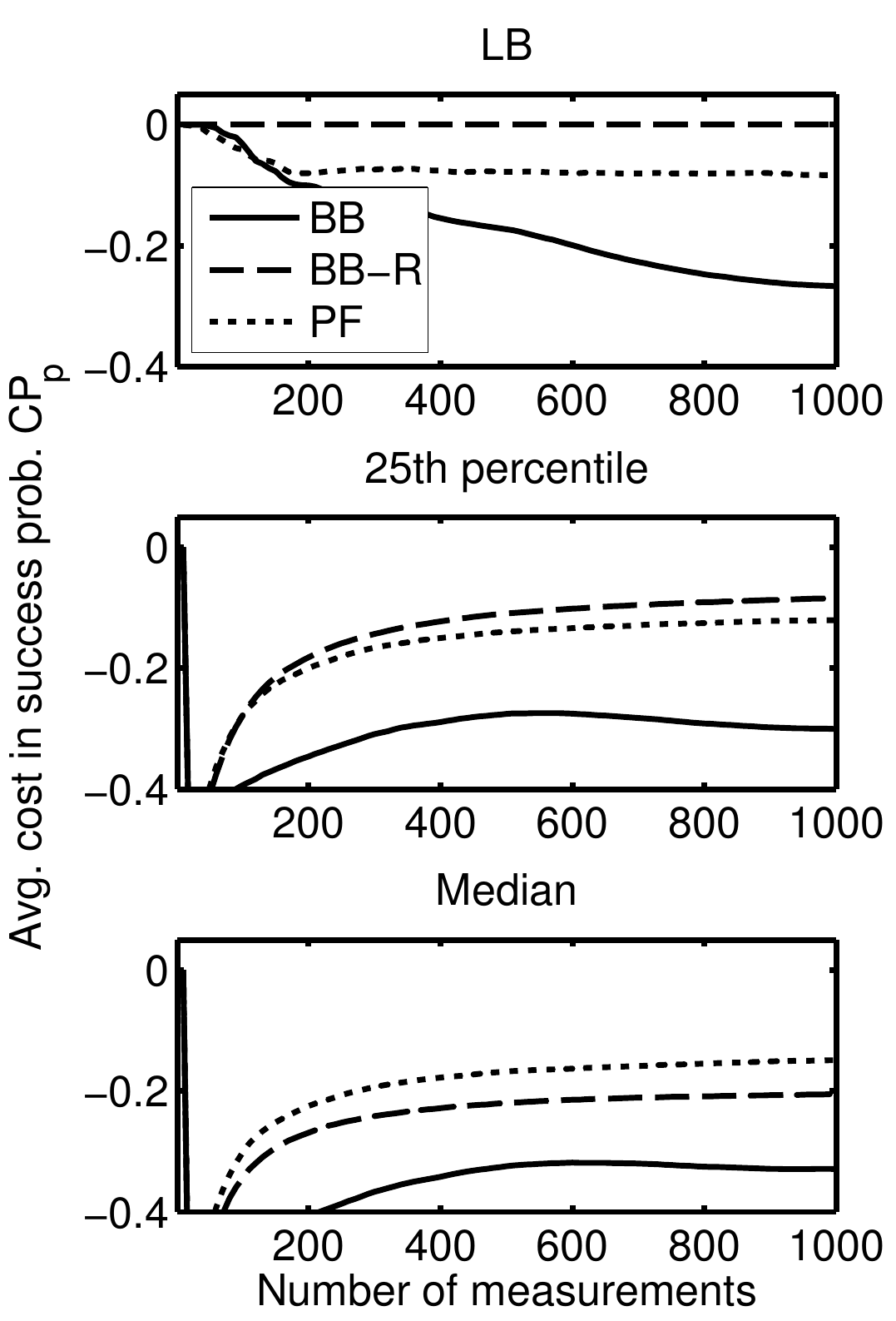}
	\caption{Cost in terms of percentage of successful transmission when estimate is greater than the PAB.  Running average of the difference between the likelihood of succesful transmission $L(z=1)$ and $\gamma=0.9$.  The difference is averaged over 72 paths and 30 simulations.  Three selection modes for estimate of PAB $\hat{y}_p(t)$: the lower bound of the confidence interval (TOP), 25th percentile of the confidence interval (MIDDLE) and median of the confidence interval (BOTTOM).\label{fig:cong_diff}}
\end{figure}

In Fig.~\ref{fig:cong_diff}, we look at the cost in percentage of successful transmission for estimates greater than the PAB.  
Whereas BB performed better in terms of cost in input rate, BB-R was the better approach in terms of cost in percentage.  PF outperformed BB in all three cases and was slightly under or over BB-R.  Combining the results from all three metrics, the PF approach provides the best balance and manages to reduce the cost input rate difference without affecting the probability of success unlike the other block-based algorithms.  In that case, increasing the input rate to the 25th percentile of the confidence interval would incur minimal cost in terms of input rate ($-5$) or percentage ($-10\%$).

\begin{figure}[!h]
	\centering
	\includegraphics[width= 0.8\linewidth]{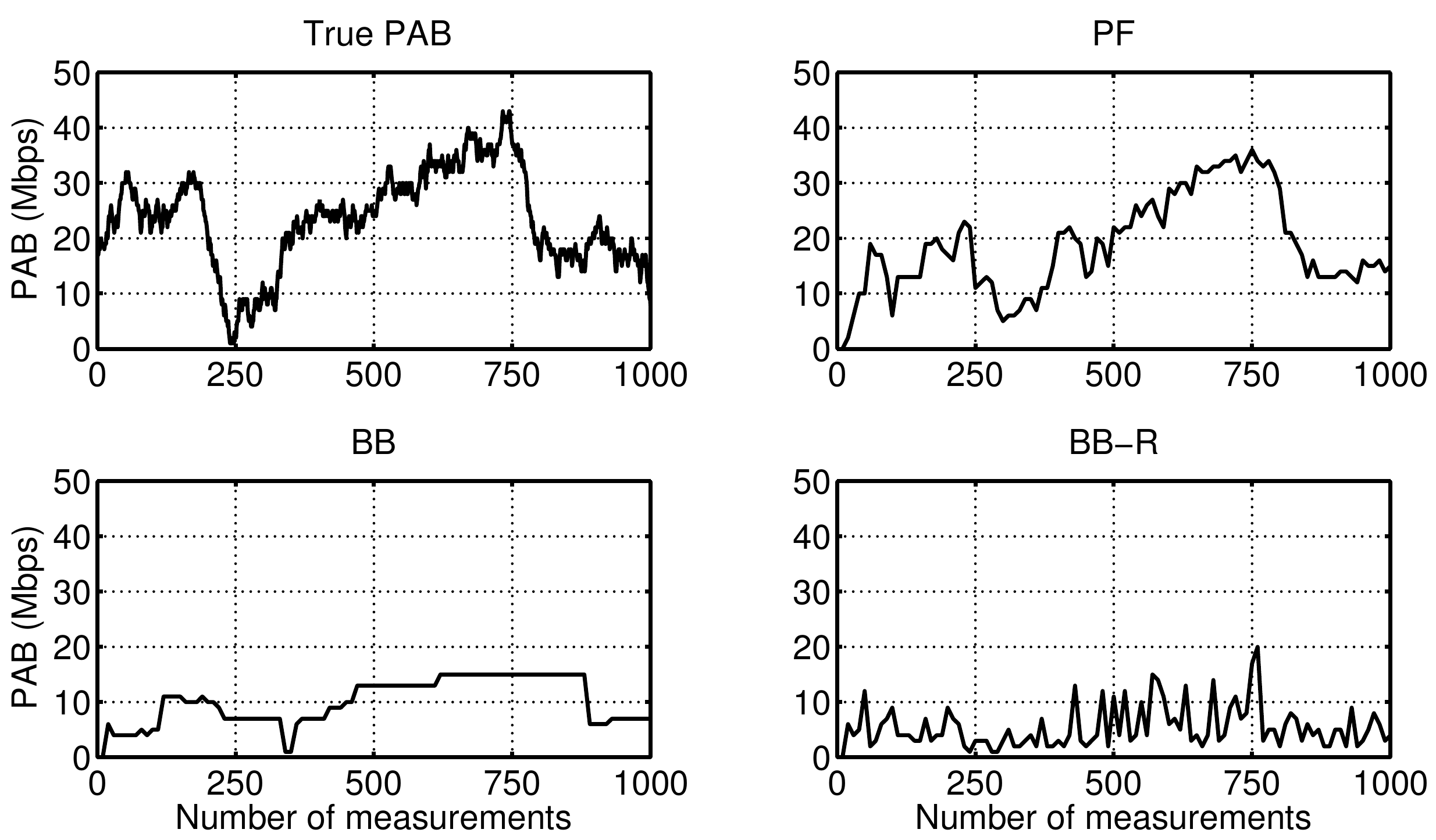}
	\caption{Tracking the PAB of a single path for 1000 iterations when the estimate is equal to the 25th percentile value of the confidence interval.\label{fig:path_track_sample}}
\end{figure}

In Fig.~\ref{fig:path_track_sample}, we show an example of tracking for a single path when the estimate is at the 25th percentile of the confidence interval.  The PF approach provides a much closer estimate than the two block-based approaches.  We also note that the PF estimate becomes more accurate after approximately 400 measurements while the other approaches are unaffected by time.  In the future, we intend to adjust our approach such that fewer measurements are required before the tracking reaches an acceptable level of accuracy.
\subsection{Online Experiments}

We implemented our tracking algorithm into a tool that we deployed on the PlanetLab network.  For our experiments, we test our approach on a topology that consists of 8 nodes\footnote{planetlab-01.cs.princeton.edu, planetlab3.csail.mit.edu, planetlab2.rutgers.edu, pl2.csl.utoronto.ca, planetlabone.ccs.neu.edu, planetlab2.csg.uzh.ch, planetlab1.millennium.berkeley.edu, planet1.zib.de}, 56 paths and 119 logical links.  We use the exact same parameter settings as in the simulations except that we set $\gamma=0.8$.  Our testing procedure consists of running the tracking algorithm for 300 measurements and probing some of the paths at regular intervals based on our estimated PAB (each experiment lasts approximately 20 minutes).  More specifically, we construct the smallest subset of paths that includes every link at least once (in this case, the test set includes 36 of the 56 paths).  After every sequence of 10 measurements, we send a constant-rate train of 500 packets on 3 of the paths from the test set at an input rate equal to our estimate of the PAB for that path.  We then calculate the average output rate and the binary outcome $z = \mathbf{1}(r'_p \geq r_p - \epsilon)$.  We observe the probability of success $\Pr(z=1)$ when the estimated PAB is chosen to be 1) the lower bound, 2) the 25th percentile value and 3) the median of the confidence interval.  We repeated the experiment five times and present the results in Tab.~\ref{tab:online}.

\begin{table}[!h]
	\caption{PlanetLab results.  Probability of success and median overestimate in case of failure for three estimated PABs.\label{tab:online}}
	\centering
	\begin{tabular}{l | c c c}
		& LB & 25th perc. & Median\\
		\hline
		$\Pr(z=1)$& $0.97 \pm 0.02$ & $0.84 \pm 0.08$ & $0.7 \pm 0.1$\\
		$OE$ (Mbps) & $10 \pm 7$ & $3 \pm 2$ & $5\pm2$\\
		$\Pr(OE > 10)$ & $0.013 \pm 0.009$ & $0.03 \pm 0.03$ & $0.05 \pm 0.02$\\		
	\end{tabular}
\end{table}

The probability of successful transmission, $\Pr(z=1)$ is averaged over the 90 tests in each experiment.  As we observed in the simulations, there is a deterioration in accuracy when we choose the PAB estimate at the median of our confidence interval.  However, for the lower bound and 25th percentile, we satisfy our constraint that $\Pr(z=1)$ should be larger than $\gamma=0.8$.  We also looked at the running average of the probability of success in time, but there was very little variation.  We noticed that, in all experiments, a small subset of paths was responsible for a large fraction of the failed tests.  As future work, we intend to investigate this problem and identify the characteristics of these paths that make them harder to estimate accurately using our method.  In case of failure, we can calculate the overestimation margin (OE); the difference between the estimate (probed rate) and the largest input rate that would satisfy our constraint: $\hat{y_p} - (r'_p + \epsilon)$.  The median OE for lower bound estimates is much larger than for the other selection methods due to the fact that there were very few overestimates.  In the other case, the over estimation is generally pretty low; under 5 Mbps.  We also look at the probability of an overestimation larger than $10$ Mbps.  In all cases, that probability was under $5\%$.

\section{Conclusion}
\label{sec:conclusion}

In this paper, we attacked the problem of tracking network-wide probabilistic available bandwidth for multiple paths.  We proposed a methodology based on chirps, belief propagation and particle filtering that we implemented in a practical and lightweight tool.  According to our simulations, our approach can track PAB with more accuracy at the same cost than block-based approaches.  We also deployed the tool in a practical environment and showed through online experimentation that it can produce estimates that satisfy our constraints in terms of probability of successful transmission (output rate almost as large as the input rate).  In the future, we intend to simulate our tool in an environment where the PABs can undergo large sudden changes or where there is a mismatch between the PAB evolution and our tracking model.

\bibliographystyle{IEEEtran}
\bibliography{availbw}

\end{document}